\documentclass[twocolumn,aps,graphicx,prb]{revtex4}
 
\usepackage{graphicx}
\usepackage{amssymb}
\usepackage{amsfonts}
\usepackage{amsmath}

\newcommand{\ket}[1]{| #1 \rangle}

\newcommand{\bra}[1]{\mbox{$\langle #1|$}}

\newcommand{\alp}{\alpha^\prime}

\begin{document}

\title{Noiseless Linear Amplification and Distillation of Entanglement
}
\author{G. Y. Xiang$^1$, T. C. Ralph$^2$, A. P. Lund$^{1,2}$, N. Walk$^2$ and G. J. Pryde$^{1,\star}$\\
$^1$\textit{Centre for Quantum Computer Technology, Centre for Quantum Dynamics,
Griffith University, Brisbane 4111, QLD, Australia.}\\
$^2$\textit{Department of Physics,
University of Queensland, Brisbane 4072, QLD, Australia.}\\
\textit{$^\star$email: g.pryde@griffith.edu.au}}

\maketitle

\textbf{The idea of signal amplification is ubiquitous in the control of physical systems, and the ultimate performance limit of amplifiers is set by quantum physics. Increasing the amplitude of an unknown quantum optical field, or more generally any harmonic oscillator state, must introduce noise\cite{CAV81}. This linear 
amplification noise prevents the perfect copying of the quantum state\cite{WOO82}, enforces
quantum limits on communications and metrology\cite{BAC04}, and is the physical mechanism that prevents the increase of entanglement via local operations. It is known that non-deterministic versions of ideal cloning\cite{DUA98} and local entanglement increase (distillation)\cite{BEN96} are allowed, suggesting the possibility of non-deterministic \textit{noiseless} linear amplification. Here we introduce, and experimentally demonstrate, such a noiseless linear amplifier for continuous-variables states of the optical field, and use it to demonstrate entanglement distillation of field-mode entanglement. This simple but powerful circuit can form the basis of practical devices for enhancing quantum technologies. The idea of noiseless amplification unifies approaches to cloning and distillation, and will find applications in quantum metrology and communications.}

 A quantum-noise-free amplifier, if it could be constructed, could aid a wide variety of quantum-enhanced information protocols, primarily through its ability to distill and purify continuous-variable entanglement. This type of entanglement is characterized by nonclassical correlations between the field quadrature, or position and momentum, variables of two or more subsystems. Such correlations represent a nonlocal resource for quantum information protocols such as continuous-variable teleportation\cite{FUR98}, dense coding\cite{BRA99,RAL02} and quantum key distribution\cite{GRO02,Silberhorn2002}. The ability to distill and purify entanglement is essential for increasing the range of these protocols. Additionally, this type of field-mode entanglement is the basis for many approaches to quantum-enhanced metrology\cite{Dowling2008}.

It is known to be impossible to perform deterministic, noiseless linear amplification. We therefore consider a device that performs the transformation
\begin{eqnarray}
 | \alpha \rangle \langle \alpha| \to \rho(\alpha) = P | g \alpha \rangle \langle g \alpha| + (1-P) | 0 \rangle \langle 0|.
\label{NLA}
\end{eqnarray}
where $g$ is a real number obeying $|g|>1$ and $ |\alpha \rangle$ is a coherent state of the field or oscillator with complex amplitude $\alpha$. We assume a heralding signal identifies which term in the output density operator has been produced by any particular run of the device. Thus, with probability $P$, noiseless amplification of the input is achieved. Without loss of generality, we assume that when amplification fails the output state is the vacuum (this can be achieved with a triggered shutter, for instance). The linearity of quantum mechanics requires that the distinguishability of two quantum states cannot be increased by any transformation. 

Consider the input states $| 0 \rangle$ and $| \alpha \rangle$. We require
\begin{eqnarray}
|\langle 0| \alpha \rangle|^2 & \le & \langle 0| \rho (\alpha)| 0 \rangle = P  |\langle 0| g \alpha \rangle|^2 +1-P
\label{NLA1}
\end{eqnarray}
The values of the overlaps are $|\langle 0| \alpha' \rangle|^2 = e^{-|\alpha'|^2}$ 
, and thus $P \le (1-e^{-|\alpha|^2})/(1-e^{-|g \alpha|^2})$. We conclude that provided $P$ is bounded in this way, non-deterministic noiseless linear amplification is physically allowed. We now present a heralded optical scheme that approximately realizes Equation (\ref{NLA}).

\begin{figure*}
\includegraphics*[width=1.5\columnwidth]{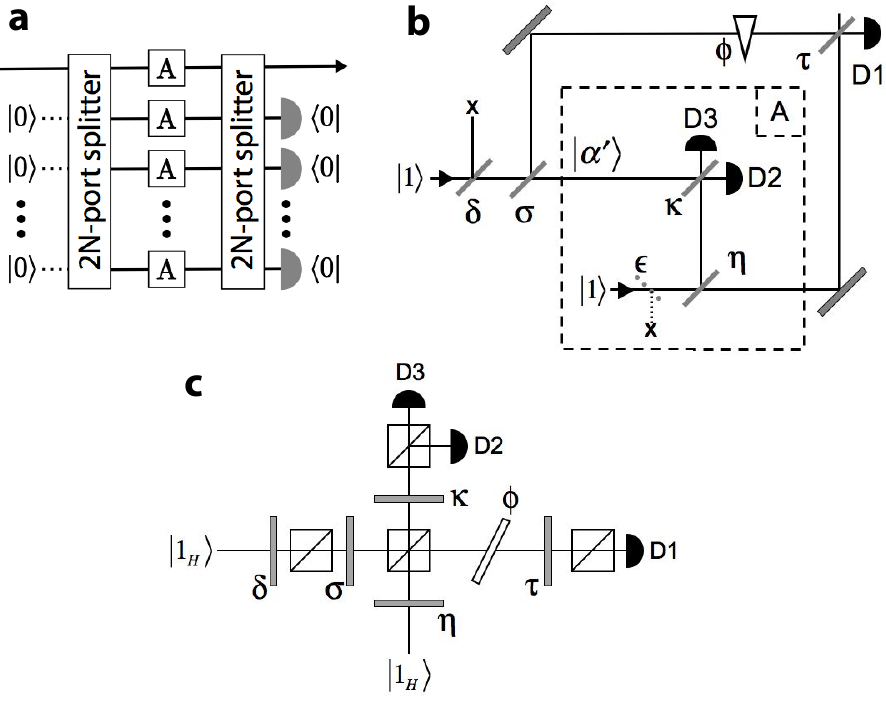}
\caption{\textbf{Design and realization of the noiseless linear amplifier.} \textbf{a.} Schematic of the full amplifier. The $2N$-port splitter evenly divides the input beam into $N$ paths. The second $2N$-port splitter coherently recombines the beams, with success if no light exits through the other ports, as determined by photon counters. The interaction labelled ``A'' is an amplifier stage. \textbf{b.} Conceptual diagram of the quantum circuit for testing an amplifier stage, shown in the dashed box marked ``A''. The amplifier stage is embedded in an interferometer (with phase shifter $\phi$) so that its phase noise properties can be tested. The roles of the various tunable beam splitters ($\delta,\sigma,\eta,\tau$) are explained in the text. $\kappa$ is usually set to the $50:50$ condition, except during calibration (see Supplementary Information). Inteference of the input with an ancilla photon, and photodetection, produces the amplification, with success heralded by either a single photon detection at D2 (and none at D3) or vice versa. The dotted beam splitter ($\epsilon$) in the ancilla input models loss in that mode (see Supplementary Information)   \textbf{c.} Experimental realization of the circuit in \textbf{b}, using polarization modes. Half wave plates and polarizing beam splitter cubes implement the tunable beam splitters, and a tilted quarter wave plate implements the phase shifter. The ancilla is a single photon, and the input state is generated from a single photon.}
%\label{fig1}
\end{figure*}

The circuit for realizing noiseless linear amplification (NLA) is shown schematically in Fig.~\textbf{1a}. The optical mode to be amplified is divided evenly between $N$ paths using a $2N$-port beam splitter. Each path then undergoes an amplification stage (Fig.~\textbf{1b}), which implements a generalization of the quantum scissors of Pegg et al.\cite{PEG98} using  
 a single photon ancilla and photon counting. The amplification is successful if exactly one photon is counted at exactly one of the conditioning detectors. The $N$ paths are then recombined interferometrically with another $2N$-port splitter. In the absence of the conditional amplifier stages ``A'', all the input light would emerge in the original mode. For the amplification scheme, successful operation of the device is heralded when photon counters on the other $N-1$ output modes register no counts, given that each amplifier stage ``A'' also yielded a heralding signal.

We first calculate the effect of this device on an input coherent state, $|\alpha \rangle$. The $2N$-port splitter divides the coherent state into the product state $|\alpha' \rangle |\alpha' \rangle |\alpha' \rangle...$, with $\alpha' = \alpha/\sqrt{N}$. Hence we can consider the effect of each amplifier stage separately. The action of the generalized quantum scissor is to truncate the coherent state to first order and simultaneously amplify it. Specifically, detection of a single photon at output port D2 and zero photons at output port D3, or detection of a single photon at output port D3 and zero photons at output port D2, produces the transformation  
\begin{eqnarray}
 |\alpha' \rangle_{a' }\to e^{-{{|\alpha'|^2}\over{2}}}\sqrt{{{\eta}\over{2}}}(1\pm \sqrt{{{1-\eta}\over{\eta}}} \hat a^{\dagger} \alpha') |0 \rangle 
\label{A1}
\end{eqnarray}
where the plus (minus) sign corresponds to the former (latter) case. In the latter case, the phase flip can be corrected by feedforward to a phase shifter. In the original quantum scissors, $\eta = 0.5$ and the truncated state is not amplified\cite{PEG98}. A sucessful coherent recombination of the modes at the second $2N$-port splitter produces
\begin{eqnarray}
e^{-{{|\alpha|^2}\over{2}}}\eta^{{{N}\over{2}}}(1 + \sqrt{{{1-\eta}\over{\eta}}} \hat a^{\dagger} {{\alpha}\over{N}})^N |0 \rangle
\label{A4}
\end{eqnarray}
In the limit of large $N$ (i.e.\ $N \gg g|\alpha|$),
\begin{eqnarray}
\lim_{N\to\infty} (1 + g \hat a^{\dagger} {{\alpha}\over{N}})^N |0 \rangle = e^{g \hat a^{\dagger} {\alpha}}|0 \rangle
\label{A5}
\end{eqnarray}
where $g =  \sqrt{(1-\eta)/\eta}$. We recognize the RHS of Equation~(\ref{A5}) as being proportional to a coherent state with amplitude $|g\alpha|$. Thus, in the large $N$ limit, the device of Fig.~\textbf{1} effects the transformation
\begin{eqnarray}
|\alpha \rangle \to \eta^{{{N}\over{2}}} e^{-{{(1-g^2)|\alpha|^2}\over{2}}} | g \; \alpha \rangle
\label{A6}
\end{eqnarray}
For $\eta < 1/2$ we have $g>1$, and hence we achieve noiseless linear amplification as per Equation~(\ref{NLA}). The quantity $g$ is the amplitude gain and $g^2$ is the intensity gain. The probability of success is given by the norm of the state, $P = \eta^{N} e^{-(1-g^2)|\alpha|^2} $,
which is state dependent and also decreases with increasing $N$. This indicates that the cost of a better approximation to the amplified state is reduced probability of success. 

The key component of the noiseless linear amplifier of Fig.~\textbf{1a} is the single \textit{amplifier stage} shown schematically in Fig.~\textbf{1b}, which we experimentally implemented using linear optics and photon counting.
Theoretically, an amplifier stage is intended to work with an input state $\ket{\alpha^\prime}$, where $|g\alpha^\prime| \ll 1$. Any ensemble of coherent states with small enough amplitude will also be linearly amplified, and within this regime this single stage of the NLA should act as desired for the full device. The amplification stage should also work independent of the phase, that is, for a state of unknown phase. Thus we can test our device using a uniform incoherent mixture, over all phases, of an ensemble of coherent states of fixed amplitude, where this amplitude can be varied for different runs of the experiment. 

We generated states of the form $\rho_{\textrm{\begin{scriptsize}in\end{scriptsize}}} = (1-k) \ket{0}\bra{0}+k\ket{1}\bra{1}$, where the attenuation is chosen such that $k=|\alp|^2$, by attenuating one arm of a weak, polarization-unentangled spontaneous parametric downconversion (SPDC) source. Such a state $\rho_{\textrm{\begin{scriptsize}in\end{scriptsize}}}$ has a theoretical fidelity of $>0.9998$ with the desired mixed coherent state when the average photon number is 0.02 in both cases. We used the other photon from the SPDC pair as the ancilla photon that drives the amplifier stage. This photon plays a dual role of heralding both the presence of the mixed coherent state input and the success of the NLA.

The experimental setup for an amplifier stage is shown in Fig.~\textbf{1c}. The beam splitters of Fig.~\textbf{1b} are implemented with half wave plates (HWPs) and polarizing beam splitter cubes, allowing tunablity of $\eta$ and the other splitting ratios. We implement an additional splitting on the input (controlled by the $\delta$ HWP setting) to prepare $\rho_{\textrm{\begin{scriptsize}in\end{scriptsize}}}$. The entire amplification stage is embedded in an interferometer, which is implemented using polarization modes, with a tuneable birefringent phase shifter $\phi$. By altering $\sigma$ and $\tau$, the amplifier stage can be operated with interferometric analysis (with tuneable interferometer beam splitting ratios) or without it. 

In order to verify that amplification has occurred, we used photon counting (see Methods Summary) to compare the measured average photon number at the input and output of the amplifier stage. 
Table~1 shows the measured intensity gain, $g^2$, as a function of the gain control reflectivity $\eta$, compared with the theoretical values, when $|\alp|^2 = 0.02$.

\begin{figure}
\begin{center}
\includegraphics*[width=\columnwidth]{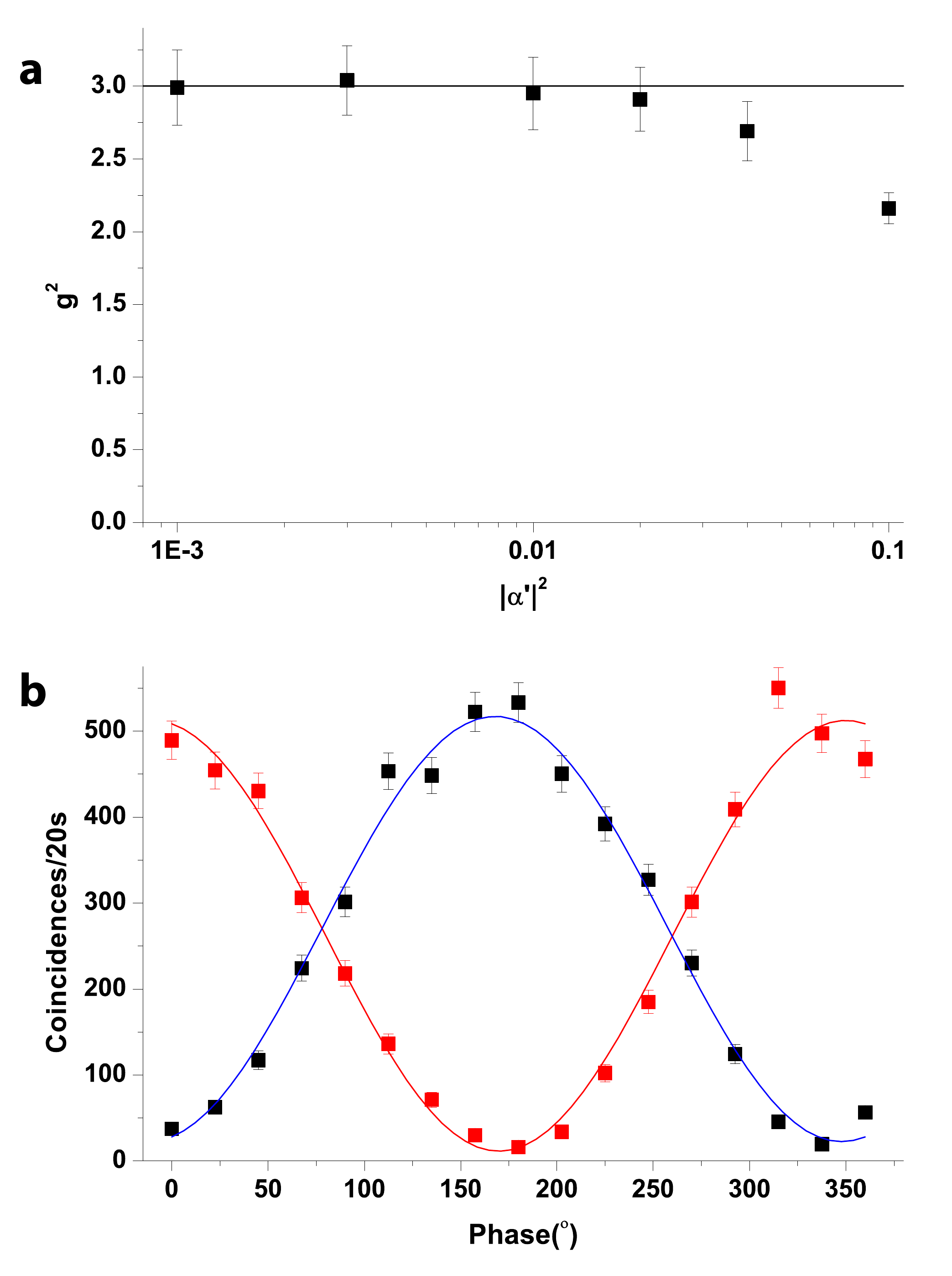}
\caption{\textbf{Gain and coherence measurements for an amplifier stage.} \textbf{a.} Linearity of the amplifier stage. For low input photon number $|\alp|^2$, the output is linear, as shown by the flat measured gain values (squares). For larger inputs, there is progressive failure of the condition $|g \alp| \ll 1$ on which the stage design is based. The quantity $(1-\eta)/\eta=3$ is shown by the horizontal line. \textbf{b.} High-visibility fringes resulting from interference of a reference beam with the amplified input state ($|\alp|^2=0.02$, $g^2 \approx 4$). The phase of the fringe depends on which conditioning detector fires, with a phase difference of $180^\circ$ between the two cases. The squares represent measured data, and the curves are least-squares fits.}
%\label{fig_data}
\end{center}
\end{figure}

The values for the gain agree well with the theoretically expected values. The fact that the gain decreases slightly---compared with the expected value---as $\eta$ decreases is in accord with a simple theoretical model incorporating preparation inefficiency of the single photon ancilla (see Supplementary Information). We checked the linearity of the amplifier's output, for a gain setting of $g^2=3$, over a range of input sizes $|\alpha^\prime|^2$ spanning two orders of magnitude, as shown in Fig.~\textbf{2a}. At low photon numbers, we have confirmed the linearity of the amplifier. At larger photon numbers, the gain begins to decrease. This is a combination of the effect of ancilla photon preparation efficiency (mentioned earlier) and a progressive failure of the condition $|g \alp| \ll 1$. Thus the amplifier stage amplifies as expected. In contrast, it is simply impossible for a deterministic amplifier to produce linear gain for this range of input states, because of the unknown phase. We emphasize that there is no pre-existing interferometric phase relation between the outputs of the SPDC (see Supplementary Information).

In order to verify that the gain process is coherent and does not add noise, we embedded the amplifier stage in an interferometer. Although our input state is mixed, we can derive a phase reference beam by splitting off part of the state at $\sigma$. (This is very similar to a ``coherent state'' that is derived from a laser beam---which is actually a mixture of coherent states of different phases\cite{Molmer1997}---where a local oscillator is usually obtained by splitting off part of the beam.) We set $\sigma=1-\eta$ such that the power in the arms of the interferometer was balanced when the amplification occurred, but not in the absence of amplification. If the conditional amplification is perfect, then no noise is added and because the two arms are balanced in power---ideally a fringe visibility of $1$ should be attainable. 
In contrast, ordinary linear amplification would introduce phase noise which would decrease the visibility significantly.
Our measured interference visibilities  are compared with theoretical values for ordinary linear amplification Table~1. The fringe corresponding to $\eta=1/5$ (nominally $g^2=4$) is shown in Fig.~\textbf{2b}. Note that the phase of the amplified light is flipped, depending on which conditioning detector fires, as predicted theoretically. We also confirmed---by varying the interferometer splitting ratio $\tau$---that the visibility is indeed \textit{optimized} when the appropriate gain is introduced (see Supplementary Information). This process is a double-check that effectively tests the amplifier stage's gain and coherence in one step.

\begin{table}
\begin{tabular}{|c|c|c|c|c|}
\hline
$\eta$ & $g^2$ (exp) & $g^2$ (thy) & $V$ (exp) & $V$ (thy, linear)  \\
\hline\hline
$1/3$ &  $2.05 \pm 0.06$ & $2$ & $0.929 \pm 0.024$ & $0.675$ \\
\hline
$1/4$ & $2.97 \pm 0.08$ & $3$ & $0.910 \pm 0.029$ & $0.514$ \\
\hline
$1/5$ & $3.85 \pm 0.10$ & $4$ & $0.936 \pm 0.022$ & $0.419$ \\
\hline
\end{tabular}
\caption{Measured gains and interferometric visibilities, for several settings of $\eta$ and for $|\alp|^2=0.02$. The measured interferometric visibilities are compared with the theoretical values for a standard linear amplifier with the same gain and input photon number, with the same interferometric configuration. The higher visibility of the experimental demonstration is evidence of low-noise coherent operation of the amplification stage, whereas a standard amplifier would add significant noise.}
%\label{exptable}
\end{table}

Our results can also be seen as an in-principle demonstration of distillation of field entanglement. We can estimate the concurrence of the field entanglement in the interferometer, with and without the amplification, using the method of Chou \textit{et al}.\cite{Chou2005}. As in that experiment, we restrict ourselves to the $\{\ket{0},\ket{1}\}$ photon number subspace for each arm. Conditional on a heralding signal the amplifier stage, the form of the state in the interferometer is $\rho = p_{00}|00 \rangle \langle00|+p_{10}|10 \rangle \langle10| + p_{01}|01 \rangle \langle 01| + d |10 \rangle \langle 01| + d^* |01 \rangle \langle10|+p_{11}|11 \rangle \langle11|$. The concurrence is given by $c=2\;\textrm{max}\; [|d|-\sqrt{p_{00}p_{11}},0]$ where $|d|=V/2$ is given by the visibility of the interferometric fringe. 

We compare the measured post-amplification concurrence ($c_{\textrm{\begin{scriptsize}out\end{scriptsize}}}$) with the theorectical concurrence ($c_{\textrm{\begin{scriptsize}in\end{scriptsize}}}$) for the interferometer input configuration---this provides the most stringent test. Setting $p_{11}=0$ and using the known values of $|\alpha'|$ and $\sigma=1-\eta=4/5$, we determine $c_{\textrm{\begin{scriptsize}in\end{scriptsize}}}=0.08$. Using the number-basis expansion of coherent states, and Equation~(\ref{A6}), it is straightforward to show that the effect of the NLA (in the large $N$ limit) on an incident number state is the transformation
\begin{eqnarray}
|n \rangle \to \eta^{{{N}\over{2}}} g^n | n \rangle
\label{E2}
\end{eqnarray}
For a state containing only zero or one photon components, this expression is true for a single amplifier stage ($N=1$). We therefore expect $p_{10}\approx p_{01}$ after the ampifier, with a commensurate increase in $|d|$. We estimate the maximum (worst-case) value of $p_{11}=(2.9 \pm 0.9) \times 10^{-4}$ from the calculated accidental coincidence rate due to 4-photon SPDC events. Together with the measured visibility $V=0.936 \pm 0.022$ and the value for $|g\alp|^2$, this leads to $c_{\textrm{\begin{scriptsize}out\end{scriptsize}}}=0.118 \pm 0.006$, which is greater than $c_{\textrm{\begin{scriptsize}in\end{scriptsize}}}$. In fact, the increase in concurrence is probably larger, as we have overestimated the contribution of $p_{11}$ to $c_{\textrm{\begin{scriptsize}out\end{scriptsize}}}$ and underestimated its contribution to $c_{\textrm{\begin{scriptsize}in\end{scriptsize}}}$. 

The principle demonstrated here directly transfers to the distillation of the more general, and widely utilized, Einstein-Podolsky-Rosen (EPR) field entanglement\cite{KIM92}. The transformation produced by the NLA operating on one arm of an EPR (or two-mode squeezed) state is given in the number basis by
\begin{eqnarray}
|EPR \rangle = K \; \sum_{n=0}^\infty \chi^n |n \rangle |n \rangle \to  K'\; \sum_{n=0}^\infty \chi^n g^n |n \rangle |n \rangle
\label{E1}
\end{eqnarray}
where the initial strength of the entanglement is given by the the parameter $\chi$, with $\chi =0$ corresponding to no entanglement and $\chi = 1$ corresponding to maximal entangement. 
We see that the transformed state still has the form of an EPR state, however the effective value of the entanglement has changed to $\chi' = g \chi$. For $|g|>1$ the entanglement has been increased, i.e. a more entangled state has been distilled. Moreover, it can be shown that the NLA can distill, and purify decohered, EPR entanglement that has experienced loss, and that high fidelity results can be obtained with only $N=2$ or $3$ (ref.\cite{RAL09}). This condition is only modestly more complex than the device demonstrated here. The ability to perform such EPR distillation would be of direct benefit to continuous-variable quantum communication protocols, by increasing the distance over which they can operate.

The concept of non-deterministic noiseless linear amplification will find applications in many fields of quantum technology. We emphasize that our demonstration of low-noise amplification is heralded, producing a freely-propagating amplified state when successful. Our device could trivially be turned into a non-deterministic, high fidelity coherent state cloner by simply dividing the output of the NLA ( for $g^2 = 2$) on a beam splitter. The fact that our device has been shown to operate well in spite of inefficient state production and heralding bodes well for more sophisticated applications. Indeed, numerical modeling (see Supplementary Information) suggests only modest improvements in efficiency would be needed for practical applications such as EPR distillation.\\

\noindent \textbf{Methods}\\
\begin{small}
\noindent \textbf{SPDC source.} 
We use a standard pulsed type-I unentangled SPDC source as in Ref.\cite{Higgins2007}. We moderate the source brightness to $<2500$ coincidences/s in order to limit the generation of $>1$ photon pair events to a negligibly small rate. 

\noindent \textbf{Amplitude measurement.} In the regime $|\alp|^2 \ll 1$, the input, ancilla and (heralded) output modes have either $0$ or $1$ photons per pulse. Thus determining the input or output average photon number corresponds to determining what proportion of pulses contain a photon, conditional on the heralding signal. 

For input calibration, we set $\eta$ and $\kappa$ such that ancilla photons travel direct to D3  with no mode splitting or interference, and set $\delta,\sigma$ and $\tau$ such that input photons travel direct to D1. $\sigma$ is then changed to its desired value (depending on which experiment we are about to perform) and $\delta$ is tuned until the coincidence efficiency in the input arm, $\mu_\textrm{in}$, reaches the desired value of $|\alp|^2=\mu_\textrm{in}=C_\textrm{\begin{scriptsize}D1D3\end{scriptsize}}/S_\textrm{\begin{scriptsize}D3\end{scriptsize}}$, equivalent to the average photon number at D1, whenever D3 fires to herald the presence of an ancilla. Here, $C$ is the coincidence rate and $S$ is the single photon count rate at the specified detector(s).  

To measure the output photon number, 
all wave plates are set to their operational values except $\tau$, which is set such that no interference occurs between the output and the reference beam---the latter is not detected. 
In this configuration, D1 measures the output signal and D3 is the detector whose output heralds successful operation of the amplifier stage. We determine $g^2 |\alp|^2=\mu_\textrm{out}=C_\textrm{\begin{scriptsize}D1D3\end{scriptsize}}/S_\textrm{\begin{scriptsize}D3\end{scriptsize}}$,  
equivalent to the average number of photons detected at D1 whenever D3 fires, heralding successful amplification. The gain is straightforwardly calculated using 
$g^2 = \mu_\textrm{out}/\mu_\textrm{in}$. 

Error bars here, and in other experimental quantities, are determined from standard error analysis techniques, and Poissonian counting statistics represent the dominant form of random errors.

\noindent \textbf{Interferometry.} We use polarization-based interferometry to measure the induced coherence between the vertically-polarized output of the amplifier and the horizontally-polarized reference beam derived from the input. The phase shifter is implemented by tilting a quarter wave plate set at its optic axis. The imperfect visibility ($V<1$) is attributed to imperfect operation of the amplifier (for example, the nonclassical interference visibility is approximately 0.96) and small instabilities of the classical interferometer. Because of small drifts in the overall circuit, we limited the measurement time for each fringe data point to 20 seconds.

Additional details on methods are provided in the Supplementary Information.

\end{small}

\vspace{15pt}

\noindent \textbf{Acknowledgements}

\noindent We thank E. H. Huntington and B. L. Higgins for useful discussions. This work was supported by the Australian Research Council.

\newpage 

\begin{widetext}

\center{\begin{large}Supplementary information for:\end{large}}\\
\center{\textit{\textbf{\begin{large}Noiseless Linear Amplification and Distillation of Entanglement\end{large}}}}\\

\center{G. Y. Xiang$^1$, T. C. Ralph$^2$, A. P. Lund$^{1,2}$, N. Walk$^2$ and G. J. Pryde$^{1,\star}$}\\
\center{$^1$\textit{Centre for Quantum Computer Technology, Centre for Quantum Dynamics,
Griffith University, Brisbane 4111, QLD, Australia.}\\
$^2$\textit{Department of Physics,
University of Queensland, Brisbane 4072, QLD, Australia.}\\
\textit{$^\star$email: g.pryde@griffith.edu.au}}
\vspace{30pt}
 \end{widetext}

\vspace{10 mm}

\noindent \textbf{Simultaneously verifying the gain and coherence properties.}

In order to simultaneously verify the gain and coherence properties of the amplifier, we measured the fringe visibility as a function of the interferometer output splitting ratio $\tau$. With $|\alp|^2=0.02$, $\sigma=0.5$ and $\eta=1/5$ (corresponding to a nominal intensity gain of 4), we observed interference fringes for various settings of $\tau$. Firstly, if the gain is close to the correct value, one expects maximum visibility at $\tau=1-\eta$, because in that condition the final beam splitter undoes the imbalance introduced by the amplifier stage. Note that the imbalance had to be introduced by the amplifier, because the input and reference beams were the same intensity. Secondly, the presence of high (ideally unit) visibility at $\tau=1-\eta$ flags coherent operation.

Fig.~\textbf{1(supplementary)} shows the results of this experiment. From the separately measured value of $g^2=3.85$, we would  expect a maximum at $\tau \approx 0.79$. As expected, we observed maximum visibility at around $\tau=1-\eta=0.8$. (We did not vary $\tau$ in smaller steps because, due to the size of the error bars and small gradient of the curve, this would not have provided any additional significant information.) The visibility at $\tau=0.8$ was measured to be $>0.9$, significantly higher than the value predicted for a standard linear amplifier at this gain (see Table 1 of the main paper).

\begin{figure}[!h]
\includegraphics*[width=\columnwidth]{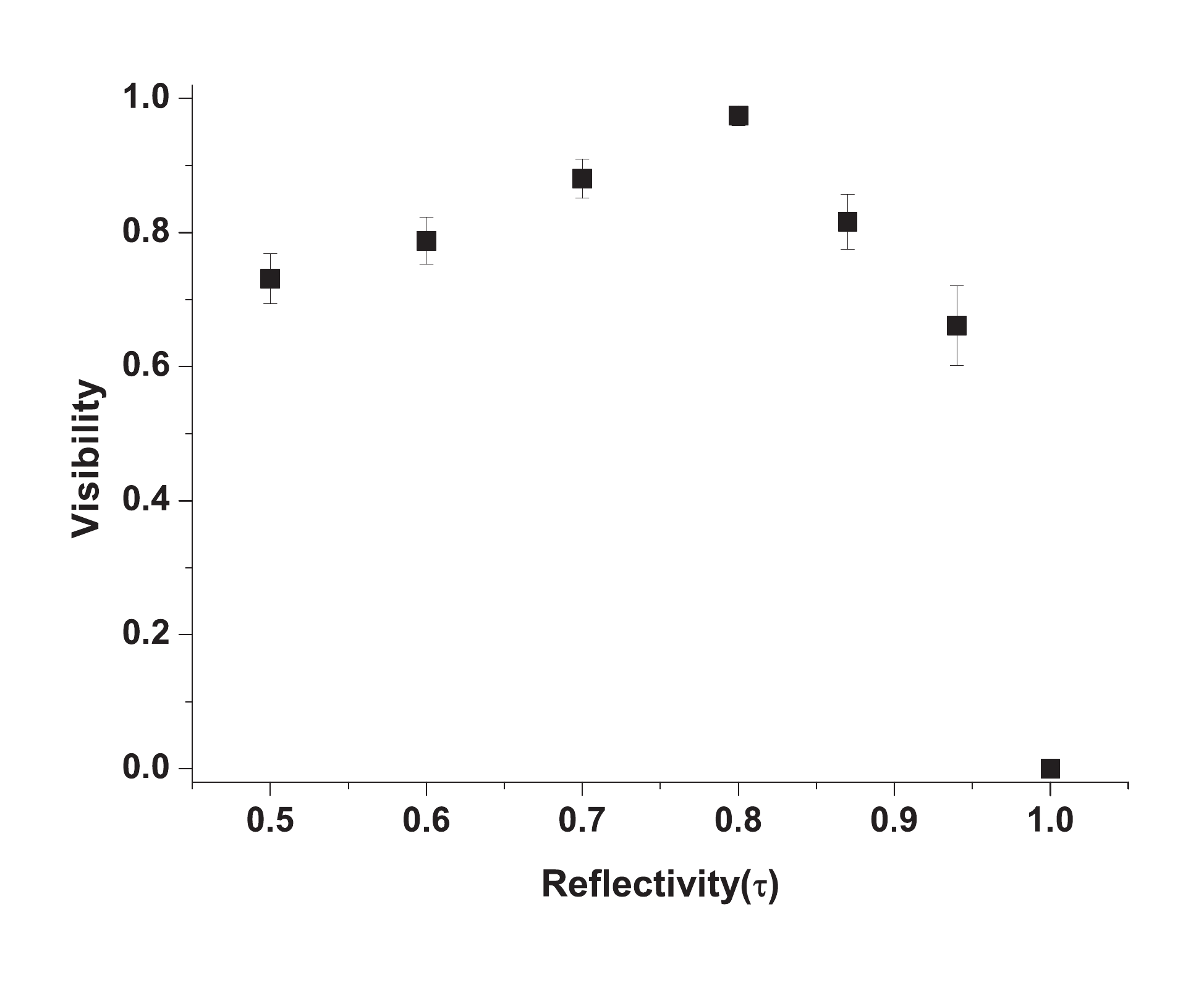}
\caption{\textbf{(supplementary)}. Measured interference visibility as a function of $\tau$ for the case $|\alp|^2=0.02$ and $\eta=1/5$. Maximum visibility is observed at $\tau=0.8$, as expected, indicating that the gain is indeed correct to balance the power in each arm at the peak visbility condition. For all points, $\sigma=0.5$.}
\label{figVis}
\end{figure}

We also observed the singles counts at the output of the interferometer, when operating in balanced mode. This corresponds to the observing the output of the interferometer independent of the amplifier stage's heralding signal. We observed negligible pre-existing coherence between the horizontal and vertical modes, as shown in Fig.~\textbf{2 (supplementary)} . Attempting to fit a sinusoid to this curve yields a visibility $\sim 0.02$. We attribute the slight deviation from zero to imperfections in the polarizing beam splitter cubes, which allow a very small leakage of the input mode into the output mode in our arrangement. 

\begin{figure}[!h]
\begin{center}
\includegraphics*[width=\columnwidth]{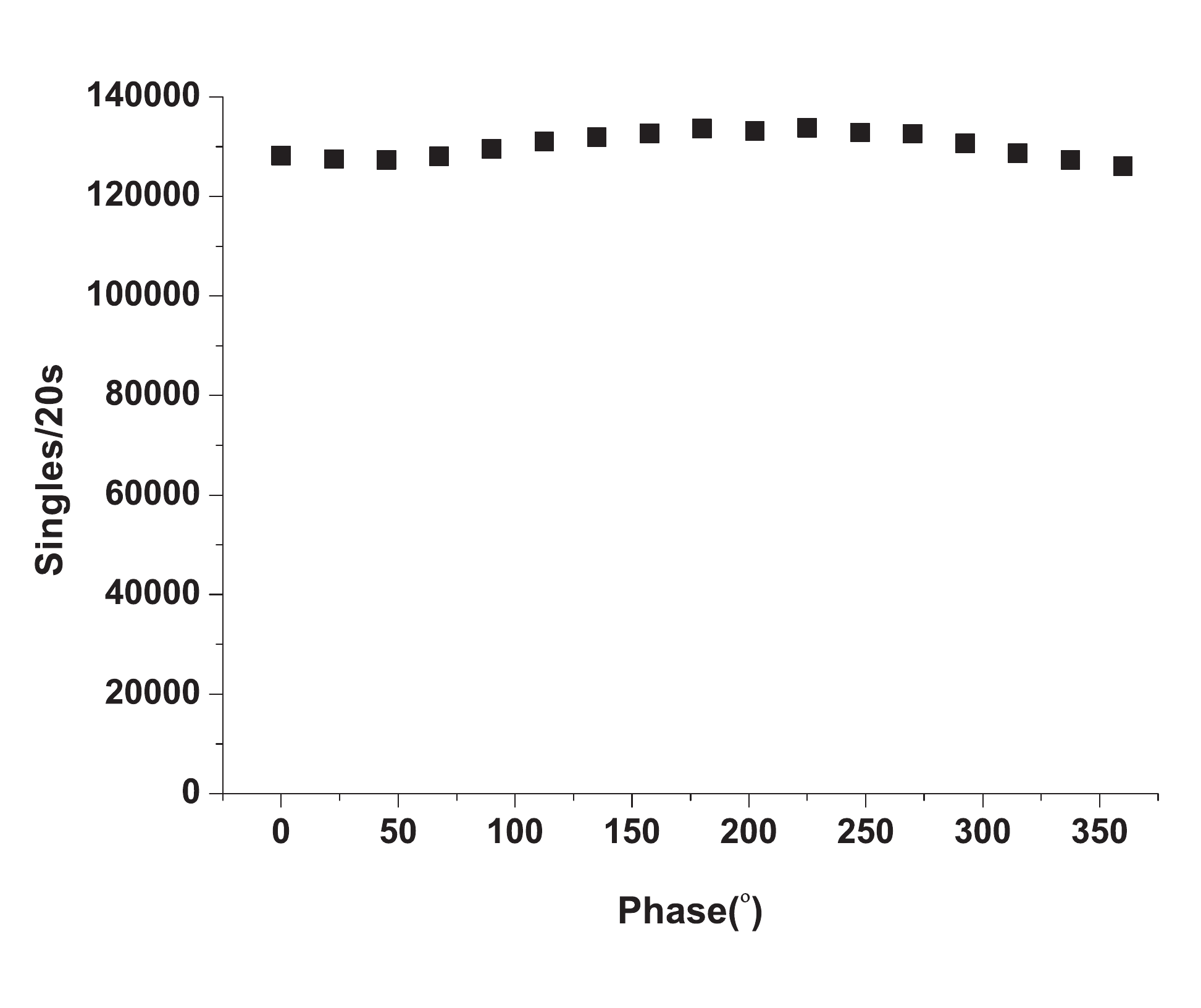}
\caption{\textbf{(supplementary).} Measured singles counts versus phase at the output of the interferometer. These data represent the measured interferometer signal \textit{without} conditioning on the amplifier stage's heralding signal. Without heralding, negligible coherence exists.}
\label{figNoCoh}
\end{center}
\end{figure}

\noindent \textbf{Loss in the ancilla mode.}

In the discussion of Table~1 of the main paper, we noted that the decrease in measured gain---relative to the expected gain---as $\eta$ decreased was a result of loss in the ancilla mode. This loss is modelled by a beam splitter in the ancilla mode, as shown in Fig.~\textbf{1b} of the main paper. We find that 
\begin{equation}
g^2_\textrm{adj}=\frac{\frac{1-\eta}{\eta}}{1+|\alp|^2 \frac{1-\epsilon}{\epsilon \eta}}.
\end{equation}
Fitting the data of table 1 to this equation yields $\epsilon=0.8 \pm 0.2$ for the preparation efficiency of the single photon (i.e.\ the probability of having had a photon present in the ancilla input whenever the amplification heralding event occurs). Notice that a key feature of this protocol is that, as long as $|\alp|^2$ is small, the gain is suprisingly insensitive to inefficiency in the ancilla input. The main effect of this inefficiency is to reduce the success probability. 

\noindent \textbf{Additional notes on calculating the gain.} 

In the Methods Summary, we discussed how the quantities $|\alp|^2$ and $|g\alp|^2$ could be derived from coincidence efficiencies. In fact, the actual photon number at the input and output deviates from these values because the detector efficiency of D1 is not unity. The detector efficiency of D1 is a product of the interference filter transmission, fibre coupling efficiency, and quantum efficiency of the commercial single photon counting module, as the detection unit is comprised of these parts. We estimate that the detector efficiency $\mu_{\begin{scriptsize}D1\end{scriptsize}} \approx 0.29 - 0.4$ and we use the nominal value $\mu_{\begin{scriptsize}D1\end{scriptsize}}=0.333$ in this work. Thus the actual conditional average input photon number is $|\alp|^2_{\textrm{\begin{scriptsize}actual\end{scriptsize}}}=\mu_\textrm{in}/\mu_{\begin{scriptsize}D1\end{scriptsize}} \approx 3 \mu_\textrm{in}$. 

However, this does not lead to a problem in calculating the gain. Again, the actual number of photons at the output, prior to detection, is higher than the detected number, so that $(g^2 |\alp|^2)_\textrm{\begin{scriptsize}actual\end{scriptsize}}=\mu_\textrm{out}/\mu_{D1}$. However, the effect of $\mu_{D1}$ cancels out in the calculation of the actual gain:
\begin{equation}
(g^2)_\textrm{actual}=\frac{(g^2 |\alp|^2)_\textrm{actual}}{|\alp|^2_\textrm{actual}}=\frac{\mu_\textrm{out} \mu_{D1}}{\mu_\textrm{in} \mu_{D1}}=\frac{\mu_\textrm{out}}{\mu_\textrm{in}}=g^2.
\end{equation}
This is the reason for using D3 as the heralding detector and D1 as the signal detector in both calibration of the input photon number and measurement of the output photon number.

\noindent \textbf{Effect of inefficient single photon ancilla and detectors.} 

Finally we consider the effect of finite efficiencies on the operation of the NLA. It is to be expected that finite detector and/or photon source efficiency will lead to mistakes in post-selecting successful operation and hence mixing in the output state. Here we consider just source efficiency however, due to the commutabilty of loss through linear networks, the effect of detector and source efficiencies will be qualitatively equivalent.

We consider single photon sources with finite efficiency $(1-\gamma)$. On the occasions when a particular source fails to produce a photon, and the event is postselected as a success, the corresponding mode will contain vacuum. Taking into account the probability of accepting a misfire event we find that the resulting density operator for the output state is

\begin{eqnarray}
\rho &=& \sum_{n=0}^N (^N_n)(1-\gamma)^{N-n} \gamma^n {{|\alpha|^{2n}}\over{N^n}} e^{-|\alpha|^2} \eta^{N-n} \nonumber\\
&\times& (1 + g \hat a^{\dagger} {{\alpha}\over{N}})^{N-n} |0 \rangle \langle 0| (1 + g \hat a{{\alpha*}\over{N}})^{N-n} 
\label{p1}
\end{eqnarray}
Surprisingly, in the large $N$ limit, all significant terms of this density operator are equal, i.e. inefficiency of the photon sources produces negligible mixing. For finite $N$ this will not be the case and mixing will occur, however Equation~(\ref{p1}) tells us this mixing will be small provided $\gamma/(1-\gamma) << \eta/|\alpha|^2$. We note that for $\gamma < 0.3$ the  example of EPR purification with N=2,3 discussed in Ref.$^{17}$ (of the main paper) satisfies this inequality by at least an order of magnitude and so should still show high fidelity in the presence of this level of loss.

\end{document}